# Isospin dependence of nucleon-nucleon elastic cross section

Qingfeng Li,[1] Zhuxia Li,[1,2,3] and Guangjun Mao[4]
[1]China Institute of Atomic Energy, P. O. Box 275(18), Beijing 102413, People's Republic of China
[2]Institute of Theoretical Physics, Academia Sinica, P. O. Box 2735, Beijing 100080, People's Republic of China
[3]Center of Theoretical Nuclear Physics, National Laboratory of Lanzhou, Heavy Ion Accelerator, Lanzhou 730000, People's Republic of China
[4]Japan Atomic Energy Research Institute, Tokai, Naka, Ibaraki 319-1195, Japan

The in-medium neutron-proton, proton-proton (neutron-neutron) scattering cross sections ($\sigma^*_{np}$, $\sigma^*_{pp(nn)}$) are studied based on QHD-II type Lagrangian within the framework of the microscopic transport theory. The results demonstrate that, for free nucleon-nucleon scattering cross sections, the isospin dependence is dominantly caused by $\rho$ meson field. The medium correction of nucleon-nucleon scattering cross sections is also isospin dependent, $\sigma^*_{np}$ depends on the baryon density weakly and $\sigma^*_{pp(nn)}$ depends on the baryon density significantly, which is due to the different effects of the medium correction of nucleon mass and $\rho$ meson mass on $\sigma_{np}$ and $\sigma_{pp(nn)}$, respectively.

PACS number(s): 25.70.−z, 24.10.Cn

## I. INTRODUCTION

With the establishment of secondary beam facilities at many laboratories around the world, the experimental studies of collisions of nuclei with large neutron or proton excess have become available, which opens up a new field of study on the structure of nuclei far from the $\beta$ stability line and nuclear reaction and the reaction mechanism at extreme of isospin asymmetry. Recently, quite a few studies on the isospin dependence of the multifragmentation [1,2] and collective flow [3–5] in heavy ion collisions at intermediate energy have been performed both experimentally and theoretically. It has been recognized that the study of collisions of neutron(proton)-rich nuclei [6] can extract the information on the equation of state as well as the liquid-gas phase transition of asymmetric nuclear matter. In order to study the neutron(or proton)-rich nuclear collisions within a transport theory approach, the isospin dependence of the mean field and the two body scattering cross sections should be introduced. In [6] the effect of the isospin dependence of mean field on the collisions of neutron-rich nuclei was studied. Concerning the nucleon-nucleon cross section, it is already known that up to hundreds MeV the free proton-neutron cross section is about 2–3 time larger than that of proton-proton (neutron-neutron)'s [7]. However, it is not clear how the nuclear medium corrects the neutron-proton and proton-proton (or neutron-neutron) elastic cross sections. It is the purpose of this paper to investigate the isospin dependence of the in-medium elastic nucleon-nucleon cross sections within the same framework used in [8–10]. The paper is arranged as follows. In the next section we will first give a brief introduction of the model and then we will give the analytic expressions of in-medium elastic cross sections for neutron-proton and proton-proton(neutron-neutron) collisions. In Sec. III we give the numerical results for both free and in-medium nucleon-nucleon cross sections. Finally, a brief summary is given in Sec. IV.

## II. FORMALISM

We start with the QHD type effective Lagrangian, in which the interaction between nucleons is described by exchanges of $\sigma$, $\omega$, $\pi$, and $\rho$ mesons. In Ref. [11], the contribution from isovector $\rho$ meson field is neglected, while in this work the contribution of $\rho$ mesons has to be considered because it plays a major role in the isospin dependence of the interaction. The Lagrangian under consideration reads as

$$L=L_F+L_I. \tag{1}$$

Here $L_F$ is the Lagrangian density for free nucleon and meson fields,

$$L_F=\bar{\Psi}[i\gamma_\mu\partial^\mu-M_N]\Psi+\frac{1}{2}\partial_\mu\sigma\partial^\mu\sigma-\frac{1}{4}F_{\mu\nu}\cdot F^{\mu\nu}+\frac{1}{2}\partial_\mu\boldsymbol{\pi}\partial^\mu\boldsymbol{\pi}-\frac{1}{4}L_{\mu\nu}\cdot L^{\mu\nu}-U(\sigma)+U(\omega)-U(\boldsymbol{\pi})+U(\boldsymbol{\rho}), \tag{2}$$

where

$$F_{\mu\nu}\equiv\partial_\mu\omega_\mu-\partial_\nu\omega_\mu \tag{3}$$

and

$$L_{\mu\nu}\equiv\partial_\mu\boldsymbol{\rho}_\nu-\partial_\nu\boldsymbol{\rho}_\mu. \tag{4}$$

$U(\sigma)$, $U(\omega)$, $U(\pi)$, and $U(\rho)$ are the self-interaction parts of the $\sigma$, $\omega$, $\pi$, and $\rho$ meson fields and the respective expressions are

$$U(\sigma)=\frac{1}{2}m_\sigma^2\sigma^2+\frac{1}{3}b\sigma^3+\frac{1}{4}c\sigma^4, \tag{5}$$

$$U(\omega)=\frac{1}{2}m_\omega^2\omega_\mu\omega^\mu, \tag{6}$$

$$U(\pi)=\frac{1}{2}m_\pi^2\boldsymbol{\pi}^2, \tag{7}$$

$$U(\rho)=\frac{1}{2}m_\rho^2\boldsymbol{\rho}_\mu\boldsymbol{\rho}^\mu. \tag{8}$$

$L_I$ is the interaction Lagrangian density of nucleons coupled to $\sigma$, $\omega$, $\pi$, and $\rho$ mesons and reads as

$$L_I=g_\sigma\bar{\Psi}\Psi\sigma-g_\omega\bar{\Psi}\gamma_\mu\Psi\omega^\mu+g_\pi\bar{\Psi}\gamma_\mu\gamma_5\boldsymbol{\tau}\cdot\Psi\partial^\mu\boldsymbol{\pi}-\frac{1}{2}g_\rho\bar{\Psi}\gamma_\mu\boldsymbol{\tau}\cdot\Psi\boldsymbol{\rho}^\mu, \tag{9}$$

where a pseudovector coupling is adopted for pion-nucleon coupling and $g_\pi=-f_\pi/m_\pi$.

It is convenient to make use of the closed time-path Green's function technique. The nucleon Green's function in the interaction picture can be defined by

$$iG_{12}=\langle T[\exp(-i\int dxH_I(x))\Psi(1)\bar{\Psi}(2)]\rangle. \tag{10}$$

Here 1, 2 represent the 4-dimension coordinates of two states, which we define on a time loop [8]. The corresponding Dyson equation for $iG_{12}$ is

$$iG_{12}=iG_{12}^0+\int dx_3\int dx_4G_{14}^0\Sigma(4,3)iG_{32}. \tag{11}$$

The nucleon self-energy $\Sigma(4,3)$ considering up to Born term reads as

$$\Sigma(4,3)=\Sigma_{\rm HF}(4,3)+\Sigma_{\rm Born}(4,3), \tag{12}$$

where the Hartree-Fock term $\Sigma_{\rm HF}(4,3)$ contributes to the mean field and the Born term $\Sigma_{\rm Born}(4,3)$ contributes to the binary collision part. After making a Wigner transformation on both sides of equation of motion for $G^{-+}$ and then adopting the semiclassical and quasiparticle approximation, we finally get the self-consistent RBUU equation for the nucleon's distribution function [8],

$$\Big\{[\partial_x^\mu-\Sigma_{\rm HF}^{\mu\nu}(x,p,\tau)\partial_\nu^p-\partial_p^\nu\Sigma_F^\mu(x,p,\tau)\partial_\nu^x]p_\mu+m^*\Big[\partial_\nu^x\Sigma_{\rm HF}^S(x,p,\tau)\partial_p^\nu-\partial_p^\nu\sum_F^S(x,p,\tau)\partial_\nu^x\Big]\Big\}\frac{f(x,p,\tau)}{E^*}=C(x,p,\tau). \tag{13}$$

Here the argument $\tau$ in mean field $\Sigma$, distribution function $f$, and collision part $C$ corresponds to the third component of isospin of nucleons ($\frac{1}{2}$ for proton and $-\frac{1}{2}$ for neutron). In the left-hand side of Eq. (13), the mean-field part

$$\Sigma_{\rm HF}^{\mu\nu}(x,p,\tau)=\partial_x^\mu[\Sigma_{H(\omega,\rho)}^\nu(x,\tau)+{\rm Re}\,\Sigma_F^\nu(x,p,\tau)]-\partial_x^\nu[\Sigma_{H(\omega,\rho)}^\mu(x,\tau)+{\rm Re}\,\Sigma_F^\mu(x,p,\tau)], \tag{14}$$

and

$$\Sigma_{\rm HF}^S(x,p,\tau)=\Sigma_{H(\sigma)}(x,\tau)+{\rm Re}\,\Sigma_F^S(x,p,\tau), \tag{15}$$

where the $\Sigma_H^\mu(x,\tau)$ and $\Sigma_H^S(x,\tau)$ are the vector and scalar components of the Hartree term of the self-energy part. $\Sigma_F^\mu(x,p,\tau)$ and $\Sigma_F^S(x,p,\tau)$ are the vector and scalar part of the Fock term contributed from the exchange of $\sigma$, $\omega$, $\pi$, and $\rho$ mesons. The Hartree terms contributed from the exchange of $\sigma$, $\omega$, and $\rho$ mesons read as

$$\Sigma_{H(\sigma)}(x,\tau)=-\frac{g_\sigma^2}{m_\sigma^2}\frac{2}{(2\pi)^3}\int d^3\mathbf{p}\frac{M^*}{(\mathbf{p}^2+M^{*2})^{1/2}}\Big[f\Big(x,p,\frac{1}{2}\Big)+f\Big(x,p,-\frac{1}{2}\Big)\Big], \tag{16}$$

$$\Sigma_{H(\omega)}^\mu(x,\tau)=\frac{g_\omega^2}{m_\omega^2}\frac{2}{(2\pi)^3}\int d^3\mathbf{p}\frac{p^\mu}{(\mathbf{p}^2+M^{*2})^{1/2}}\Big[f\Big(x,p,\frac{1}{2}\Big)+f\Big(x,p,-\frac{1}{2}\Big)\Big], \tag{17}$$

$$\Sigma_{H(\rho^0)}^\mu(x,\tau)=\begin{cases}\dfrac{g_\rho^2}{m_\rho^2}\dfrac{1}{2(2\pi)^3}\displaystyle\int d^3\mathbf{p}\frac{p^\mu}{(\mathbf{p}^2+M^{*2})^{1/2}}\Big[f\Big(x,p,\frac{1}{2}\Big)-f\Big(x,p,-\frac{1}{2}\Big)\Big], & \tau=\dfrac{1}{2},\\ \dfrac{g_\rho^2}{m_\rho^2}\dfrac{1}{2(2\pi)^3}\displaystyle\int d^3\mathbf{p}\frac{p^\mu}{(\mathbf{p}^2+M^{*2})^{1/2}}\Big[f\Big(x,p,-\frac{1}{2}\Big)-f\Big(x,p,\frac{1}{2}\Big)\Big], & \tau=-\dfrac{1}{2}.\end{cases} \tag{18}$$

Here $M^*$ denotes the effective mass of nucleon. Because $\pi$ is a pseudoscalar meson there is no contribution from $\pi$ meson to the Hartree term and for $\rho$ meson, only the third component of $\rho$ meson field contributes to the Hartree term since there is no charge exchange at the Hartree level. The Hartree terms given in Eqs. (16), (17), and (18) should be solved self-consistently with the field equations of $\sigma$, $\omega$, and $\rho$ mesons. The Fock terms are usually neglected in the transport theory approach for simplicity.

The right-hand side of Eq. (13) is the collision term derived from the same Lagrangian as that of mean field, which we will stress in this paper,

$$C(x,p,\tau)=C_{\rm el}(x,p,\tau)+C_{\rm in}(x,p,\tau). \tag{19}$$

Here we only calculate the elastic part, which can be written as

$$C_{\rm el}(x,p,\tau)=\frac{1}{2}\int\frac{d^3\mathbf{p}_2}{(2\pi)^3}\int\frac{d^3\mathbf{p}_3}{(2\pi)^3}\int\frac{d^3\mathbf{p}_4}{(2\pi)^3}\times(2\pi)^4\delta^4(p+p_2-p_3-p_4)\times W_{\rm el}(p,p_2,p_3,p_4)[F_2-F_1], \tag{20}$$

where $W_{\rm el}(p,p_2,p_3,p_4)$ is the transition probability. The relation between the transition probability and differential cross section is

$$\int\ v\sigma_{\rm el}d\Omega=\int\frac{d^3\mathbf{p}_3}{(2\pi)^3}\int\frac{d^3\mathbf{p}_4}{(2\pi)^3}(2\pi)^4\delta^4(p+p_2-p_3-p_4)\times W_{el}(p,p_2,p_3,p_4). \tag{21}$$

Inserting Eq. (21) into Eq. (20) we obtain

$$C_{\rm el}(x,p,\tau)=\frac{1}{2}\int\frac{d^3\mathbf{p}_2}{(2\pi)^3}d\Omega\,\sigma_{\rm el}v[F_2-F_1], \tag{22}$$

where $v$ is the Møller velocity and $\sigma_{el}$ is the elastic differential cross section. $F_1$ and $F_2$ are the Pauli block factors, which can be written as

$$F_1=f_1(x,p,\tau)f_2(x,p,\tau)[1-f_3(x,p,\tau)][1-f_4(x,p,\tau)], \tag{23}$$

$$F_2=f_3(x,p,\tau)f_4(x,p,\tau)[1-f_1(x,p,\tau)][1-f_2(x,p,\tau)]. \tag{24}$$

The detailed expressions of the proton-proton, neutron-neutron, and neutron-proton cross sections are given as follows:

$$\frac{d\sigma_{pp}}{d\Omega}=\frac{d\sigma_{nn}}{d\Omega}=\sum_{i=1}^{10}\frac{A_i}{2}[D_i(s,t)+E_i(s,t,u)+(s,t\to u)], \tag{25}$$

$$\frac{d\sigma_{np}}{d\Omega}=\sum_{i=1}^{10}\frac{A_i}{2}[d_iD_i(s,t)+e_iE_i(s,t,u)+(s,t\to u)], \tag{26}$$

where $i$ denotes the contributions from individual $\sigma,\omega,\pi,\rho$ exchange and crossing terms. The coefficients $A_{i=1,10}$ are

$$A_1=\frac{1}{(2\pi)^2s}g_\sigma^4,\quad A_2=\frac{1}{(2\pi)^2s}g_\omega^4,$$

$$A_3=\frac{1}{(2\pi)^2s}\left(\frac{M^*}{M_N}g_\pi\right)^4,\quad A_4=\frac{1}{(2\pi)^2s}\left(\frac{g_\rho}{2}\right)^4,$$

$$A_5=\frac{1}{(2\pi)^2s}g_\sigma^2g_\omega^2,\quad A_6=\frac{1}{(2\pi)^2s}g_\sigma^2\left(\frac{M^*}{M_N}g_\pi\right)^2,$$

$$A_7=\frac{1}{(2\pi)^2s}g_\sigma^2\left(\frac{g_\rho}{2}\right)^2,\quad A_8=\frac{1}{(2\pi)^2s}g_\omega^2\left(\frac{M^*}{M_N}g_\pi\right)^2,$$

$$A_9=\frac{1}{(2\pi)^2s}g_\omega^2\left(\frac{g_\rho}{2}\right)^2,\quad A_{10}=\frac{1}{(2\pi)^2s}\left(\frac{M^*}{M_N}g_\pi\right)^2\left(\frac{g_\rho}{2}\right)^2. \tag{27}$$

The coefficients $d_i$, $e_i$, in Eq. (26) are given as

$$d_1=d_2=d_5=1,\quad d_3=d_4=5,\quad d_6=d_8=d_{10}=0,$$

$$d_7=d_9=-1,$$

$$e_1=e_2=e_5=0,\quad e_3=e_4=e_{10}=-4,\quad e_6=e_7=e_8=e_9=2, \tag{28}$$

and the expressions of $D_i$, $E_i$, read as

$$D_1=\frac{(t-4M^{*2})^2}{2(t-m_\sigma^2)^2},\quad E_1=-\frac{t(t+s)+4M^{*2}(s-t)}{8(t-m_\sigma^2)(u-m_\sigma^2)},$$

$$D_2=\frac{2s^2+2st+t^2-8M^{*2}s+8M^{*4}}{(t-m_\omega^2)^2},$$

$$E_2=\frac{(s-2M^{*2})(s-6M^{*2})}{2(t-m_\omega^2)(u-m_\omega^2)},$$

$$D_3=\frac{t^2}{2(t-m_\pi^2)^2},\quad E_3=\frac{(4M^{*2}-s-t)t}{8(t-m_\pi^2)(u-m_\pi^2)},$$

$$D_4=D_2(m_\omega\to m_\rho),\quad E_4=E_2(m_\omega\to m_\rho),$$

$$D_5=-\frac{4(2s+t-4M^{*2})M^{*2}}{(t-m_\sigma^2)(t-m_\omega^2)},$$

$$E_5=\frac{t^2-4M^{*2}s-10M^{*2}t+24M^{*4}}{4(t-m_\sigma^2)(u-m_\omega^2)}+\frac{(t+s)^2-2M^{*2}s+2M^{*2}t}{4(t-m_\omega^2)(u-m_\sigma^2)},$$

$$D_6=0,E_6=\frac{t^2+st-8M^{*2}t-4M^{*2}s+16M^{*4}}{8(t-m_\sigma^2)(u-m_\pi^2)}+\frac{t(t+s)}{8(t-m_\pi^2)(u-m_\sigma^2)},$$

$$D_7=D_5(m_\omega\to m_\rho),\quad E_7=E_5(m_\omega\to m_\rho),$$

$$D_8 = 0, \quad E_8 = \frac{(t+s-4M^{*2})(t+s-2M^{*2})}{4(t-m_\omega^2)(u-m_\pi^2)} + \frac{t^2-2M^{*2}t}{4(t-m_\pi^2)(u-m_\omega^2)},$$

$$D_9 = \frac{2(2s^2+2st+t^2-8M^{*2}s+8M^{*4})}{(t-m_\omega^2)(t-m_\rho^2)},$$

$$E_9 = \frac{(s-2M^{*2})(s-6M^{*2})}{2}\left(\frac{1}{(t-m_\omega^2)(u-m_\rho^2)} + \frac{1}{(t-m_\rho^2)(u-m_\omega^2)}\right),$$

$$D_{10} = 0, \quad E_{10} = E_8(m_\omega \to m_\rho). \tag{29}$$

$s$, $t$, and $u$ are Mandelstain variables defined as

$$s = (p+p_2)^2 = (p_3+p_4)^2, \tag{30}$$

$$t = (p-p_3)^2 = (p_2-p_4)^2, \tag{31}$$

$$u = (p-p_4)^2 = (p_2-p_3)^2 = 4M^{*2}-s-t. \tag{32}$$

## III. NUMERICAL RESULTS

Before coming to the calculation of the $\sigma_{np,pp(nn)}$, we have to make some preparations. First, we introduce the common used form factor of nucleon-meson-nucleon vertex, which reads as

$$F_i(t) = \frac{\Lambda_i^2}{\Lambda_i^2 - t}, \tag{33}$$

where the subscripts $i$ represent the different meson exchange vertex and $\Lambda_i$ is the cutoff mass for meson i. In this work the values of $\Lambda_\sigma = 2m_\sigma$, $\Lambda_\omega = 1.1m_\omega$, $\Lambda_\pi = 700$ MeV and $\Lambda_\rho = 800$ MeV are used.

Now let us first consider the contributions from $\sigma$ and $\omega$ mesons to the $\sigma_{np,pp(nn)}$. It is well known that there is strong cancellation between the $\sigma$ and $\omega$ terms in the mean field. For nucleon-nucleon cross sections, which are beyond the mean field approximation, one can easily find that the direct cancellation between the terms corresponding to $\sigma$ and $\omega$ is lost. Therefore they become very sensitive to $g_\sigma$ and $g_\omega$. Figure 1 shows the calculated $n$-$p$ scattering cross sections from the contributions of $\sigma$ and $\omega$ field with parameter sets given in Table I comparing with the experimental data taken from [7].

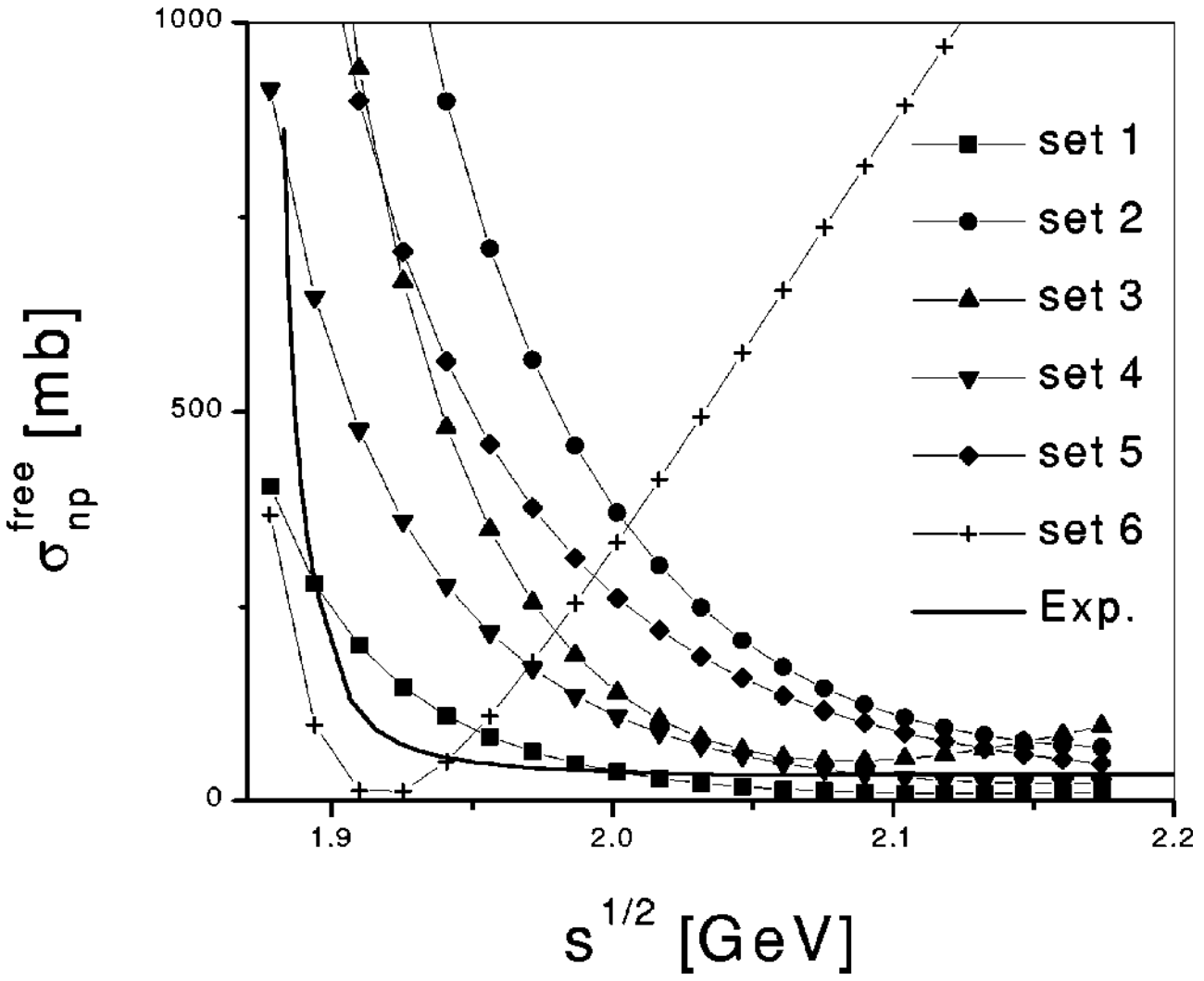

FIG. 1. The free nucleon-proton scattering cross sections where only the contributions of $\sigma$ and $\omega$ are taken into account. The parameter sets used are given in Table I. The experimental data are also given in the figure.

The parameter sets for $\sigma$ and $\omega$ given in Table I are taken from [12,11], $g_\rho$ is taken from [13] by fitting the symmetry energy of the ground state of asymmetric nuclear matter, and a common used value is taken for the coupling constant $g_\pi$. From the figure we can see that there is a marked different behavior of cross sections with different sets of coupling constants. It means that the nucleon-nucleon cross sections are indeed very sensitive to the coupling constants $g_\sigma$ and $g_\omega$. Because the main contribution to the cross section comes from $\sigma$ meson at low energy and $\omega$ meson at high energy, if $g_\omega$ is much larger than $g_\sigma$ the cross section will rise too fast with the increase of energy as shown by the curve calculated with set 6 in the figure. After studying the behavior of cross sections as a function of energy with at least 30 sets of the coupling constants taken from [12,11,14] we find if $g_\omega > g_\sigma$ and $g_\omega$-$g_\sigma <\sim 2$ the tendency of $\sigma_{np}^{\rm free}$ and $\sigma_{pp(nn)}^{\rm free}$ with increasing energy is approximately to that of experimental data. In the following calculation, the param-eter set 4 is taken because a best fit to the experimental data is obtained according to the numerical results shown in Fig. 1, which is consistent with [11]. Figure 2 shows the ratio of the calculated $\sigma_{np}^{\text{free}}$ to $\sigma_{pp(nn)}^{\text{free}}$ as a function of $s^{1/2}$ as well as that of the corresponding experimental ones. We can see that the ratio between the calculated $\sigma_{np}^{\text{free}}$ and $\sigma_{pp(nn)}^{\text{free}}$ contributed only from $\sigma$ and $\omega$ mesons is about unity and far from the experimental ratio especially at low energy. However, $\rho$ meson plays major role in the isospin dependence of the nucleon-nucleon interaction, we then consider the contributions of $\pi$ and $\rho$ mesons. We find that the contribution from $\pi$'s is very small, only about several millibarns, and $\rho$ meson provides the dominant contribution to the isospin dependence of the nucleon-nucleon scattering cross sections. The $\rho$-$\sigma$ and $\rho$-$\omega$ crossing terms $D_7$ and $D_9$ give the largest contribution. The contribution of $D_7$ term to $\sigma_{np}$ is positive and is negative to $\sigma_{pp(nn)}$. The contribution of $D_9$ to $\sigma_{pp(nn)}$ and to $\sigma_{np}$ is opposite to $D_7$. But the magnitude of $D_7$ is larger than that of $D_9$ so the net contribution of them enhances $\sigma_{np}^{\text{free}}$ and suppresses $\sigma_{pp}^{\text{free}}$. Figures 3(a) and 3(b) show the experimental and calculation results of $\sigma_{np}^{\text{free}}$ and $\sigma_{pp(nn)}^{\text{free}}$, respectively. The general behavior of the isospin dependence of nucleon-nucleon scattering cross sections is in consistent with the experimental data, though the theoretical calculation results are a little larger than those from experiments. The parameters of Walecka model are originally fixed by the saturation properties of nuclear matter. It is not clear what values of the parameters there should be in free space. From the DBHF calcualtion of Brockmann and Machleidt [15], as well as the DDHF studies for finite nuclear system in [16,17], it was shown that $g_\sigma$ and $g_\omega$ should be density dependent. Mao *et al.* [11] proposed a density and momentum dependence of the coupling constants of the sigma and omega field. And based on it, a trend of decrease of the cross sections at low energy as the density decreasing from normal density to $0.25\rho_0$ was shown and the cross sections at $0.25\rho_0$ was quite similar to the free cross section [11]. Therefore, the calculated free nucleon-nucleon cross sections, which are larger than experimental data at low energy range, seem to be reasonable. Since this work is mainly concentrated on the isospin dependence of the nucleon-nucleon cross section we will not involve ourselves with this aspect. We have also tested the dependence of the difference between $\sigma_{np}^{\text{free}}$ and $\sigma_{pp}^{\text{free}}$ on $g_\rho$, we find the best fit to the experimental data of $\sigma_{np}^{\text{free}}$ to $\sigma_{pp}^{\text{free}}$ can be obtained when $g_\rho$ is taken to be $\sim 4$, which is inconsistent with [13], where the best fit to the symmetry energy is obtained based on mean field calculation.

TABLE I. Parameter sets. $m_\sigma = 550$, $m_\omega = 783$, $m_\pi = 138$, $m_\rho = 770$, $M_0 = 939$, $g_\pi = 1.434$, and $g_\rho = 4.23$.

| | $g_\sigma$ | $g_\omega$ | $b[\mathrm{fm}]^{-1}$ | $c$ | $m^*$ | $K$[MeV] | $E_{\rm bin}$[MeV] | $\rho_0[\mathrm{fm}]^{-3}$ |
|---|---|---|---|---|---|---|---|---|
| Set 1 | 5.93 | 6.70 | 67.98 | 523.06 | 0.85 | 400 | −16 | 0.15 |
| Set 2 | 8.94 | 9.60 | −11.19 | 25.47 | 0.75 | 300 | −16 | 0.15 |
| Set 3 | 9.40 | 10.95 | −0.69 | 40.44 | 0.70 | 380 | −15.57 | 0.145 |
| Set 4 | 6.90 | 7.54 | −40.49 | 383.07 | 0.83 | 380 | −15.76 | 0.145 |
| Set 5 | 7.24 | 7.34 | −39.14 | 696.71 | 0.85 | 200 | −15.75 | 0.1484 |
| Set 6 | 11.31 | 15.30 | −35.84 | 596.18 | 0.70 | 200 | −15.75 | 0.1484 |

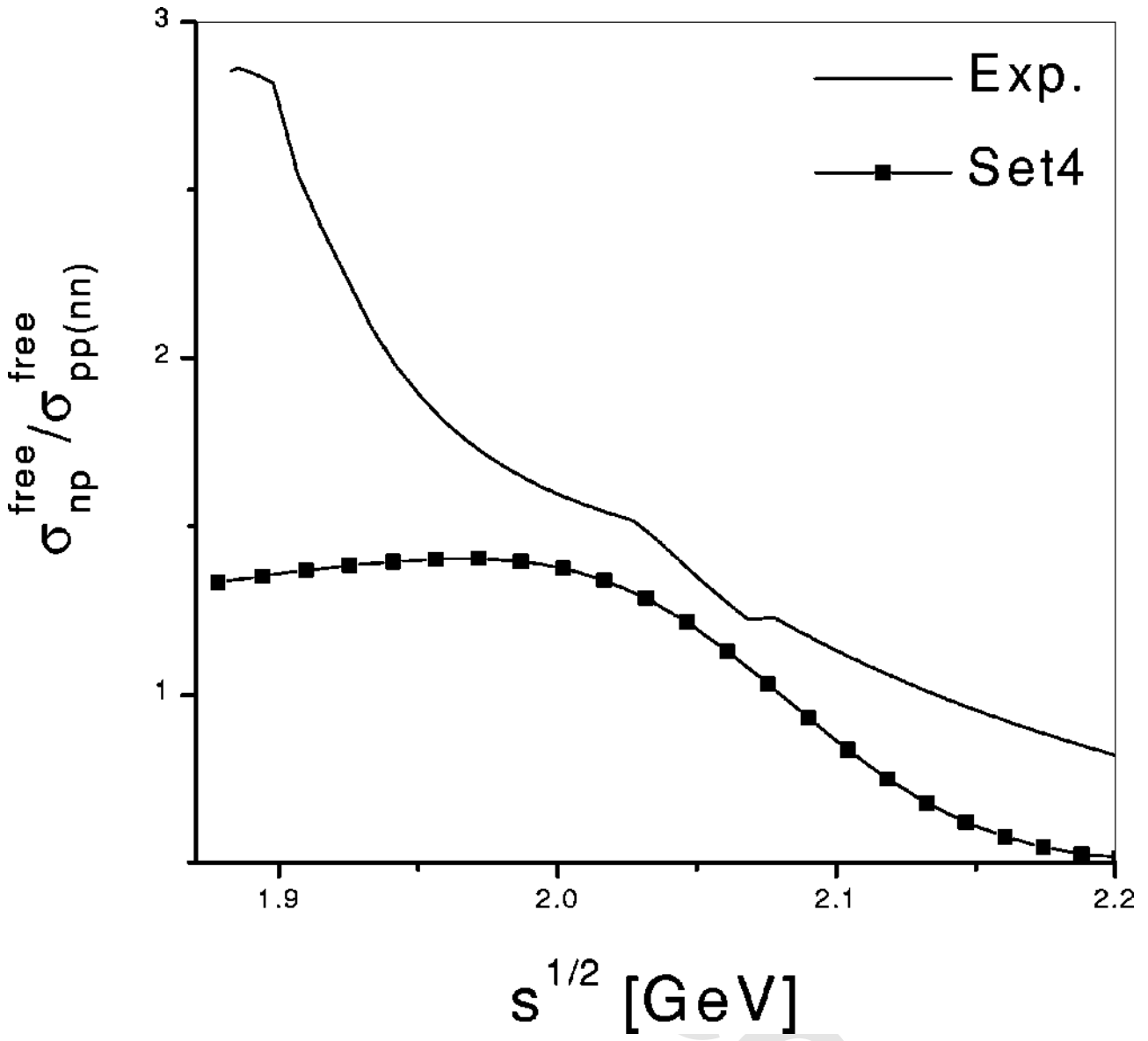

FIG. 2. The comparison of the ratio of $\sigma_{np}^{\text{free}}/\sigma_{pp(nn)}^{\text{free}}$ between the calculation results and experimental data when the contribution of $\rho$ meson field is not taken into account.

Now let us investigate the behavior of in-medium cross sections of $\sigma_{np}^*$ and $\sigma_{pp(nn)}^*$. In addition to consider the medium correction of the nucleon mass it is also very important to take the medium correction of the $\rho$ mass into account which has already attracted a lot of studies [18–20]. Here we take the medium correction of $\rho$ mass by using the Brown-Rho scaling [18], i.e.,

$$\frac{m_\rho^*}{m_\rho} = \frac{1}{1 + c\,\dfrac{\rho}{\rho_0}}, \tag{34}$$

where c is a variable parameter. In this paper we simply fit it to the experimental data given in [21] , that is, $m_\rho^*$ equals to 610 MeV when $\rho/\rho_0 = 0.7$, so $c$ is equal to 0.3747. With this value, the $\rho$ mass decreases more stiffly with increase of density than that given in [13], but the difference is not large when the density is not high. It will not change the main conclusions given in the paper. The masses of $\sigma$ and $\omega$ mesons are fixed since one usually considers the sigma and omega model as an effective one.

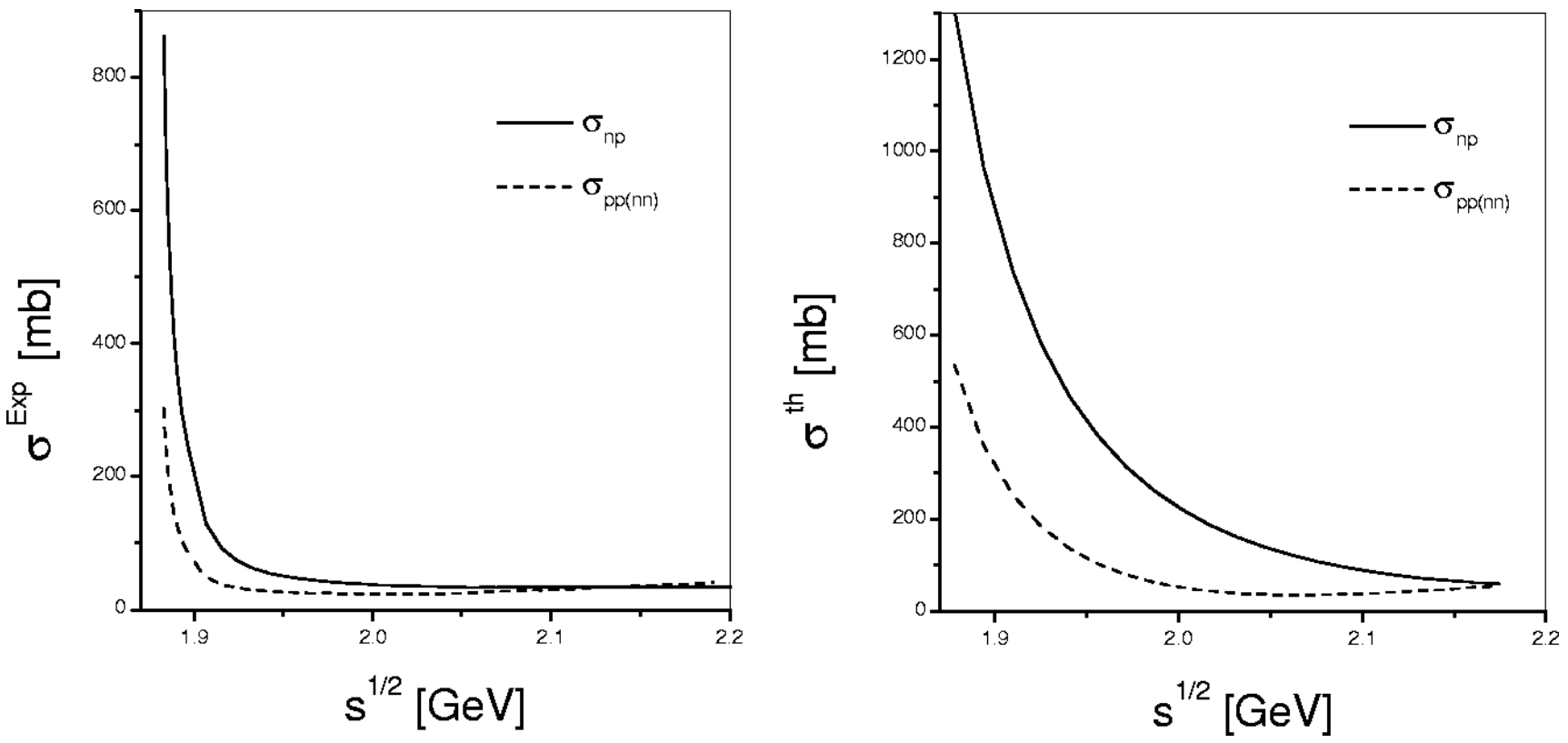

FIG. 3. The comparison between the experimental data (left) and the calculation results (right) for $\sigma_{pp(nn)}^{\text{free}}$ and $\sigma_{np}^{\text{free}}$.

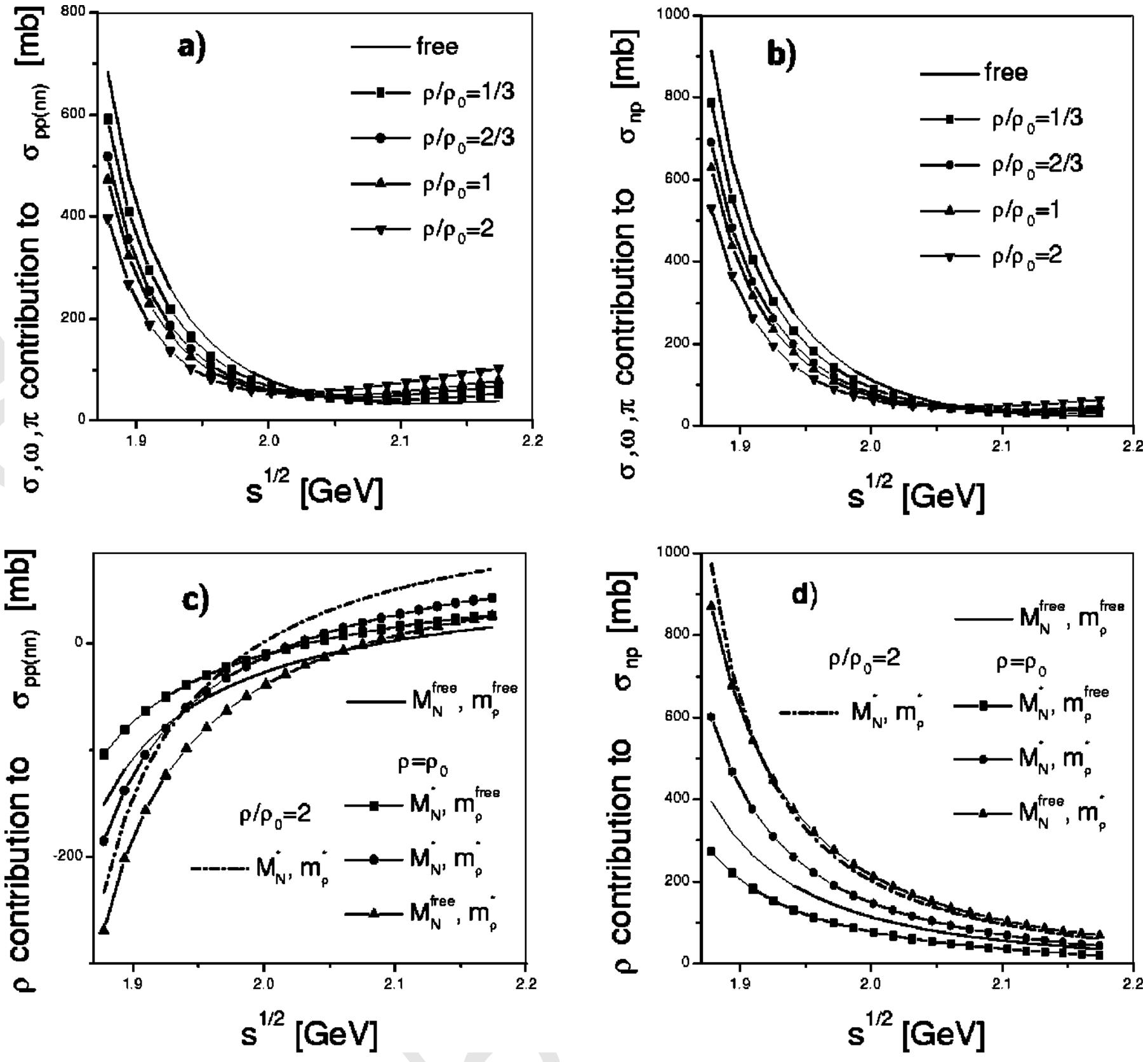

FIG. 4. The contributions of $\sigma$, $\omega$, and $\pi$ to (a) $\sigma^*_{pp(nn)}$ and (b) $\sigma^*_{np}$ at density $\rho/\rho_0=0$, 1/3, 2/3, 1, and 2, respectively, as well as the contribution of $\rho$ meson field to (c) $\sigma^*_{pp(nn)}$ and (d) $\sigma^*_{np}$ at density $\rho/\rho_0=0$, 1, 2. The contribution of $\rho$ meson to $\sigma^*_{pp(nn)}$ and $\sigma^*_{np}$ at normal density for three different cases, i.e., (a) only effective mass of nucleon taken into account, (b) only effective mass of $\rho$ meson taken into account, and (c) effective mass of both nucleon and $\rho$ meson taken into account are also shown (c) and (d), respectively.

In order to investigate the respective effects from different meson fields on in-medium and free nucleon-nucleon scattering cross sections, we show $\sigma^*_{pp(nn)}$ and $\sigma^*_{np}$ contributed only from $\sigma$, $\omega$, and $\pi$ mesons at baryon densities $\rho/\rho_0=0$, 1/3, 2/3, 1, and 2 in Figs. 4(a)and 4(b), respectively. From the figures we can easily find that if we only include the contributions of $\sigma$, $\omega$, and $\pi$ mesons, both $\sigma^*_{pp(nn)}$ and $\sigma^*_{np}$ will be monotonously suppressed at lower energy and then enhanced slightly at higher energy with the increase of the baryon density. In Figs. 4(c) and 4(d), we show $\sigma^*_{pp(nn)}$ and $\sigma^*_{np}$ contributed from $\rho$ meson involved terms (including the crossing terms of $\rho$ field with $\sigma$, $\omega$, $\pi$ field) at $\rho/\rho_0=0$, 1 and 2, respectively. In contrast with Figs. 4(a) and 4(b), the tendency of the contribution of $\rho$ meson to $\sigma^*_{pp(nn)}$ and $\sigma^*_{np}$ is not the same as the nuclear densidy increasing. It is shown that the contribution from $\rho$ meson field to $\sigma^*_{np}$ increases with the increase of the nucleon density while the contribution of $\rho$ meson to the $\sigma_{pp(nn)}$ decreases at lower energy and increases at higher energy as the density increases. To understand the reason, we further investigate the effects resulting from the medium correction of $\rho$ meson and nucleon mass on the $\rho$ meson contributions to $\sigma^*_{pp(nn)}$ and $\sigma^*_{np}$, respectively. In Figs. 4(c)and 4(d) we also show the $\rho$ meson contributions to $\sigma^*_{pp(nn)}$ and $\sigma^*_{np}$ at normal density for three different cases, i.e., (a) only effective mass of nucleon taken into account, (b) only effective mass of $\rho$ meson taken into account, and (c) effective mass of both nucleon and $\rho$ meson taken into account. After comparing the results for these cases, we can find that in the energy range of our investigation, the medium correction of the nucleon mass mainly leads to enhance the $\rho$ meson contribution to $\sigma^*_{pp(nn)}$ and suppress that to $\sigma^*_{np}$, while the medium correction of $\rho$ mass leads to suppress the $\rho$ meson contribution to $\sigma_{pp(nn)}$ in lower energy and enhance that to $\sigma^*_{pp(nn)}$ at higher energy but leads to enhance the $\rho$ meson contribution to $\sigma^*_{np}$ in all of the investigated energies. The net effect resulting from the medium correction of nucleon mass and $\rho$ meson mass enhanses $\sigma^*_{np}$ at all energy investigated and suppresses $\sigma^*_{pp(nn)}$ at low energy and enhances $\sigma^*_{pp(nn)}$ slightly at higher energy. Therefore, we can conclude that, for $\sigma^*_{pp(nn)}$, the medium effects from the contributions of the $\sigma$ and $\omega$ meson terms are in the same direction with that of $\rho$ meson term, while for $\sigma^*_{np}$, the effects from the contributions of the $\sigma$ and $\omega$ meson terms are opposite to that of $\rho$ meson term, which result in a strong density dependence of $\sigma^*_{pp(nn)}$ and a weak density dependence of $\sigma^*_{np}$ as can be seen in Figs. 5(a) and 5(b). Here we show the in-medium cross sections $\sigma^*_{np}$ and $\sigma^*_{pp(nn)}$ at one and 2 times normal density, respectively, in which all of the contributions of $\sigma$, $\omega$, $\pi$, and $\rho$ mesons and the medium corrections of both nucleon and $\rho$ mass are included.

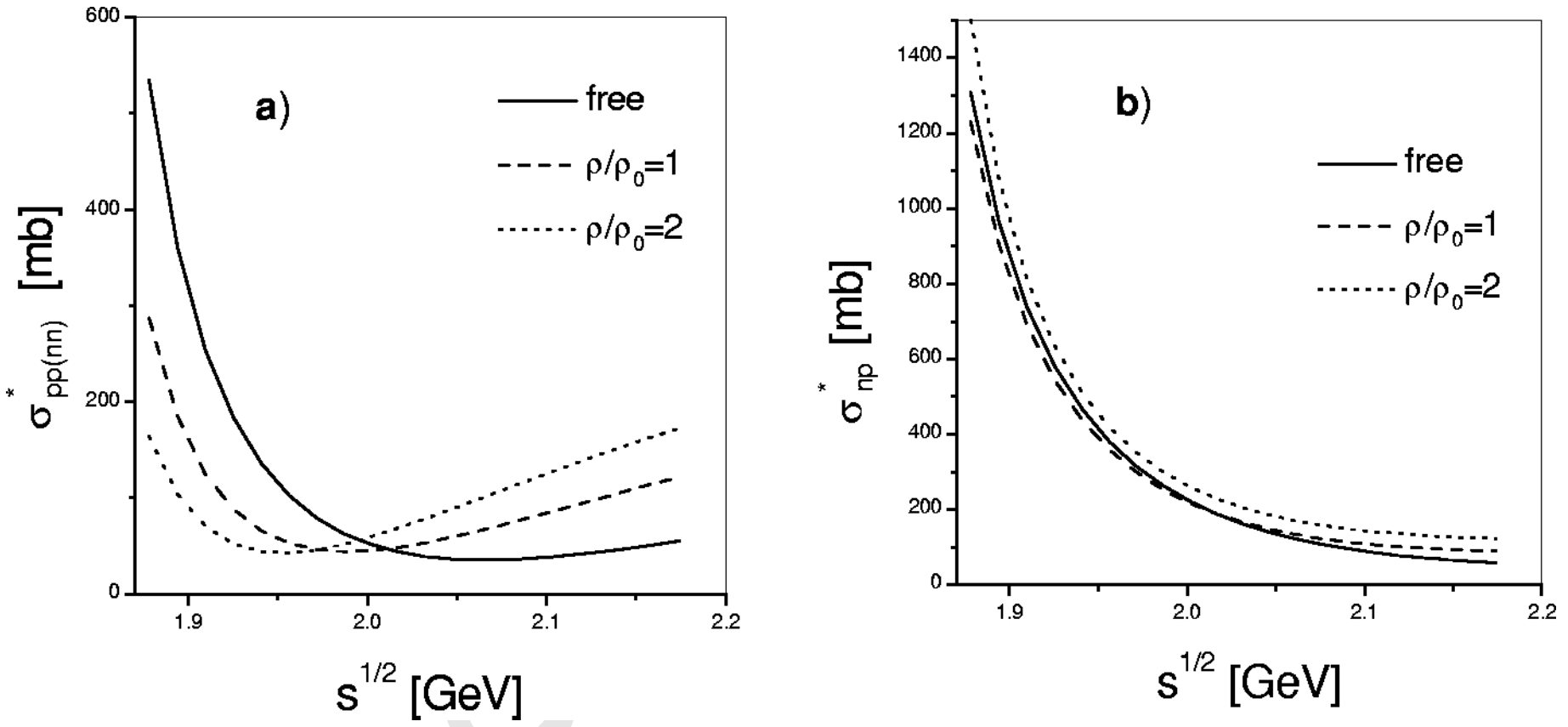

FIG. 5. The in-medium cross sections (a) $\sigma^*_{pp(nn)}$ and (b) $\sigma^*_{np}$ at $\rho/\rho_0 = 0,\ 1,\ 2.$

## IV. SUMMARY AND OUTLOOK

In this paper, we first study the marked different behavior of $\sigma^{\rm free}_{pp(nn)}$ and $\sigma^{\rm free}_{np}$ when different sets of coupling constants are adopted, it seems that if $g_\omega > g_\sigma$ and $g_\omega - g_\sigma < \sim 2$ we can approximately reproduce the tendency of the energy dependence of $\sigma^{\rm free}_{pp(nn)}$ and $\sigma^{\rm free}_{np}$. Then we study the isospin dependence of the nucleon-nucleon scattering cross section and we find if only considering the contributions of $\sigma$ and $\omega$ fields, we cannot reproduce the difference of $\sigma^{\rm free}_{pp(nn)}$ and $\sigma^{\rm free}_{np}$ at lower energy which is found in experimental data. Our results indicate that $\rho$ meson field plays a dominant role in the isospin dependence of the nucleon-nucleon elastic cross sections, while $\pi$ meson field plays a negligible role. By taking the contribution of $\rho$ meson field into account, the isospin dependence of the nucleon-nucleon elastic cross sections can be reproduced reasonably well. For the medium effects, in addition to the medium correction of the nucleon mass, we stress on the effects of the medium correction of the $\rho$ mass. Our results have demonstrated the different behaviors of the effects resulting from the medium correction of $\rho$ mass and nucleon mass on $\sigma^*_{pp(nn)}$ and $\sigma^*_{np}$, respectively, which leads to enhance $\sigma^*_{np}$ and suppress $\sigma^*_{pp(nn)}$ at lower energy and enhance it at higher energy with the increase of the density. So, for $\sigma^*_{pp(nn)}$, the medium effects from the contributions of the $\sigma$ and $\omega$ meson terms are at the same direction with that of $\rho$ meson term, the total effects consequently result in a strong density dependence in $\sigma^*_{pp(nn)}$. While, for $\sigma^*_{np}$, the medium effects on the contributions of $\sigma$ and $\omega$ mesons and that of $\rho$ mesons to $\sigma^*_{np}$ are of opposite, this cancellation makes the density dependence of $\sigma^*_{np}$ very weak. Thus, a commonly adopted phenomenological expression for in-medium nucleon-nucleon elastic cross section $\sigma^* = [1 - \alpha(\rho/\rho_0)]\sigma^{\rm free}$ should make some change concerning the different behavior of medium corrections of $\sigma^*_{np}$ and $\sigma^*_{pp(nn)}$. In order to take the isospin dependence of the medium correction into account, an isospin dependent $\alpha$ should be introduced in the phenomenological expression when applying it to the transport theory approach for heavy ion collisions at low and intermediate energy.

## ACKNOWLEDGMENTS

This work was supported by the National Natural Science Foundation of China under Grant Nos. 19975073 and 19675069. G.M. acknowledges the STA foundation for financial support and the members of the Research Group for Hadron Science at Japan Atomic Energy Research Institute for their hospitality.